\documentclass[aps,prl,twocolumn,superscriptaddress,showpacs,longbibliography]{revtex4-2}
\usepackage{graphicx} 
\usepackage{amsmath} 
\usepackage{amssymb} 
\usepackage{amsfonts}
\usepackage{bbm,bm}
\usepackage{lineno}

\ifx\pdftexversion\undefined
\usepackage[dvips]{hyperref}
\else
\usepackage[colorlinks, allcolors=blue]{hyperref}
\fi
\hypersetup{
  colorlinks = true, linkcolor = blue
}

\begin{document} 

\title{Observation of pairs of atoms at opposite momenta \\ in an equilibrium interacting Bose gas}

\author{Antoine Tenart}
\affiliation{Universit\'e Paris-Saclay, Institut d'Optique Graduate School, CNRS, Laboratoire Charles Fabry, 91127, Palaiseau, France}
\author{Ga\'etan Herc\'e}
\affiliation{Universit\'e Paris-Saclay, Institut d'Optique Graduate School, CNRS, Laboratoire Charles Fabry, 91127, Palaiseau, France}
\author{Jan-Philipp Bureik}
\affiliation{Universit\'e Paris-Saclay, Institut d'Optique Graduate School, CNRS, Laboratoire Charles Fabry, 91127, Palaiseau, France}
\author{Alexandre Dareau}
\affiliation{Universit\'e Paris-Saclay, Institut d'Optique Graduate School, CNRS, Laboratoire Charles Fabry, 91127, Palaiseau, France}
\author{David Cl\'ement}
\affiliation{Universit\'e Paris-Saclay, Institut d'Optique Graduate School, CNRS, Laboratoire Charles Fabry, 91127, Palaiseau, France}

\begin{abstract}
Quantum fluctuations play a central role in the properties of quantum matter. In non-interacting ensembles, they manifest as fluctuations of non-commuting observables, quantified by Heisenberg inequalities \cite{heisenberg1927}. In the presence of interactions, additional quantum fluctuations appear, from which many-body correlations and entanglement arise \cite{amico2008entanglement}. Weak interactions are predicted to deplete Bose-Einstein condensates by the formation of correlated pairs of bosons with opposite momenta \cite{bogoliubov1947, lee1957}. Here, we report the observation of these atom pairs in the depletion of an equilibrium interacting Bose gas \cite{carcy2021}. Our measurements of atom-atom correlations, both at opposite and close-by momenta \cite{carcy2019momentum, cayla2020}, allow us to characterise the equilibrium many-body state. We also show that the atom pairs share the properties of two-mode squeezed states \cite{loudon1987, braunstein2005}, including relative number squeezing \cite{orzel2001,esteve2008squeezing, bucker2011}. Our results illustrate how interacting systems acquire non-trivial quantum correlations as a result of the interplay between quantum fluctuations and interactions \cite{pitaevskii1991}.
\end{abstract}

\maketitle

Interaction-induced quantum correlations and entanglement are essential features of many-body physics that arise from interactions. Understanding these properties poses challenges for both experimentalists and theorists \cite{amico2008entanglement}. In experiments, various methods are now able to probe correlations between individual particles in interacting systems. Therefore, quantitative tests of many-body theories become accessible in regimes where theoretical predictions exist. A conceptually simple example, which we study here, is that of weakly-interacting bosons, described by the celebrated Bogoliubov theory.

In a seminal paper \cite{bogoliubov1947}, N. Bogoliubov introduced a theoretical description of weakly-interacting bosons, motivated by the observation of superfluidity of liquid Helium. In that context, an emblematic prediction of this theory is the linear spectrum of excitations that results from interactions and has been observed in a large variety of systems \cite{miller1962, ozeri2005, fontaine2018, stepanov2019}. A second central result is a microscopic description of the many-body ground state as a macroscopic occupation of the lowest-energy state of the trap, a Bose-Einstein condensate (BEC), and a fraction of bosons outside the BEC, the quantum depletion, originating from interaction-induced quantum fluctuations \cite{pitaevskii1991}. The quantum depletion is expected to extend over a large momentum range and to feature correlations between opposite momenta ${\bm k}$/$-{\bm k}$. Measurements of the momentum density $\rho({\bm k})$ have confirmed the presence of these large momentum components \cite{sokol1995, xu2006}, and Bogoliubov's prediction of the quantum depleted fraction \cite{lopes2017}. However, the ground-state correlations at opposite momenta have not yet been observed. 

An intuitive picture of the origin of ${\bm k}$/$-{\bm k}$ correlations can be derived from Bogoliubov's theory which assumes weak interactions and considers only processes involving two atoms in the BEC \cite{bogoliubov1947} whose momenta are close to zero. To ensure momentum conservation after the interaction, the two bosons leaving the BEC (to populate the quantum depletion) have opposite momenta ${\bm k}$/$-{\bm k}$, and form a momentum-correlated pair \cite{lee1957}. Although apparently simple, this mechanism should be contrasted with non-linear processes generating similar momentum-correlated pairs in out-of-equilibrium configurations, where non-linearities efficiently drive resonant processes with both momentum and energy being conserved and for which (semi-)classical pictures hold. Out-of-equilibrium examples range from parametric down conversion (PDC) in quantum optics \cite{burnham1970} and dissociation of diatomic molecules in atomic physics \cite{greiner2005} to elastic collisions in high-energy physics \cite{arnison1982} and at ultra-low temperatures \cite{perrin2007observation}. In contrast, individuating the creation of a single ${\bm k}$/$-{\bm k}$ pair in the equilibrium many-body ground state would violate (kinetic) energy conservation, indicating a failure of classical pictures in the case of the quantum depletion. This highlights the collective nature of these two-body interactions, delocalized over the BEC wave function, as a result of quantum fluctuations.  

\begin{figure*}[ht!]
    \centering
    \includegraphics[width=1.6\columnwidth]{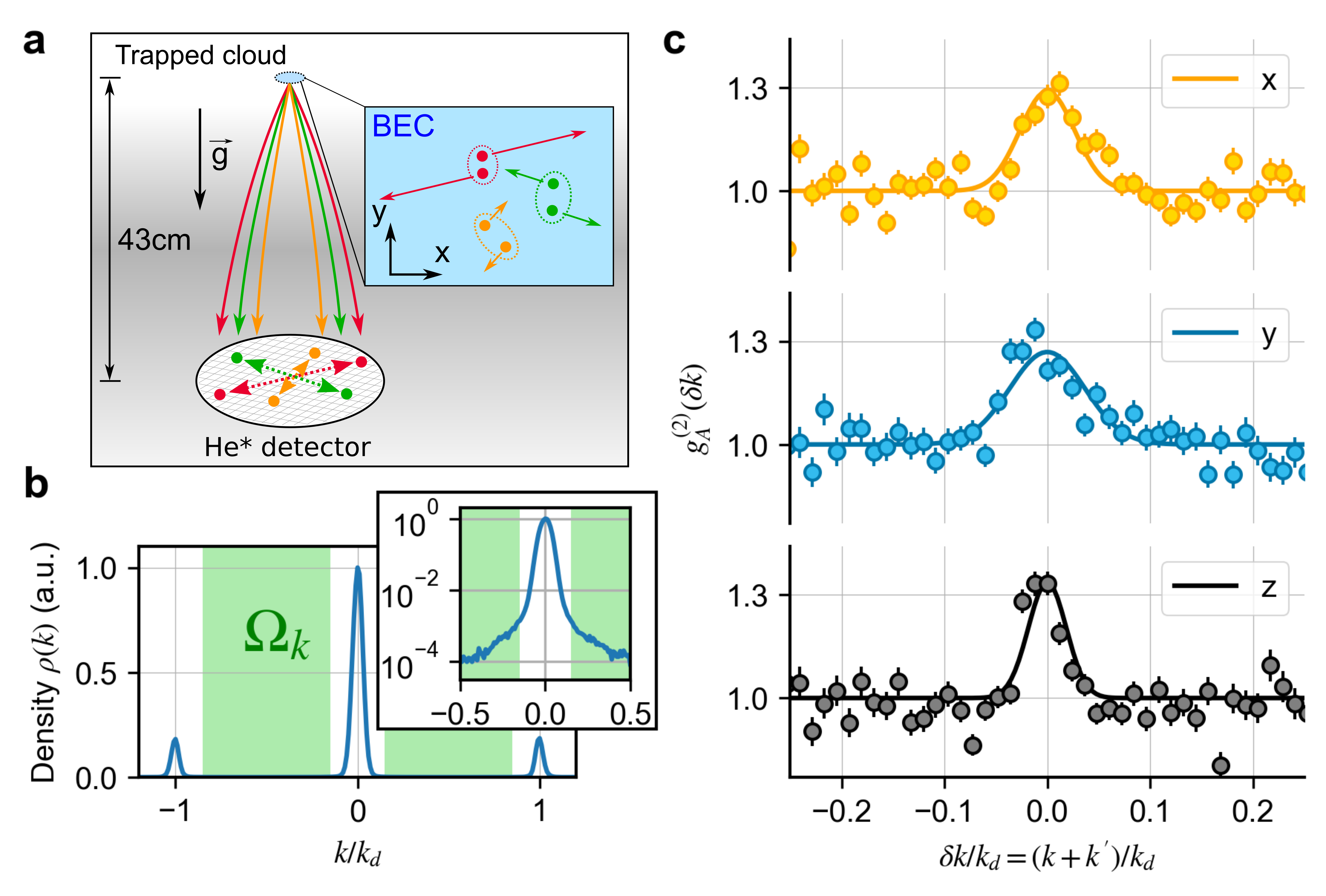}
    \caption{{\bf (a)}. Schematics of the experiment. A gas of weakly-interacting $^4$He$^*$ atoms is released from a lattice trap and undergoes a free fall towards the He$^*$ detector where atoms are detected individually. The inset depicts the many-body ground state which contains a BEC (uniform light blue color) and the quantum depletion made of pairs of atoms with opposite momenta (colored circles). When the trap is abruptly switched off, the many-body ground state is projected onto the momentum basis and the atom pairs fall onto the He$^*$ detector with opposite momenta (atoms from the BEC are not shown on the detector). {\bf (b)} 1D cut $\rho(k)$ through the experimental momentum density $\rho(\bm{k})$. The peaks correspond to the (coherent) BEC component. The depletion of the BEC, corresponding to long tails in $\rho(k)$, is visible in log-scale (inset). The volume $\Omega_{k}$ over which atom-atom correlations are computed is indicated as green shaded area.
    {\bf (c)}. Atom-atom correlations revealing pairs of atoms with opposite momenta. 1D cuts through $g_{A}^{(2)} (\delta {\bm k})$ along the axis of the 3D optical lattice. The transverse integration is $\Delta k_{\perp}=3 \times 10^{-2} \ k_d$ and the longitudinal voxel size is $\Delta k_{\parallel}=1.2 \times 10^{-2} \ k_d$. The data is fitted by Gaussian functions (solid lines). The error bars are obtained from the inverse square root of the number of counts in the voxels.
    }
    \label{Fig1}
\end{figure*}

These interaction-induced quantum fluctuations create a coherent superposition of ${\bm k}$/$-{\bm k}$ pairs \cite{lee1957} that play a central role in the equilibrium properties of interacting bosons. Though their origin differs from that of PDC photon pairs, the Hamiltonian terms of pair creation are analogous and lead to similar paired states. Therefore, we expect the correlations between the momenta ${\bm k}$ and $-{\bm k}$ to be analogous to those of two-mode squeezed states \cite{loudon1987}. Such correlations illustrate the non-trivial nature of the Bogoliubov many-body ground state as two-mode squeezed states constitute a primary resource of entanglement with continuous variables \cite{braunstein2005}. Observing these fascinating properties requires probing interacting bosons with a single-particle resolution in momentum space. To this end, our quantum-gas experiment combines a long time-of-flight (TOF) to access momentum space, and a three-dimensional detection method of individual metastable Helium-4 ($^4$He$^*$) atoms (see Fig.~\ref{Fig1}a). This approach \cite{cayla2018single}  allows us to seek atom pairs through the correlator $\langle \hat{a}^{\dagger}({\bm k}) \hat{a}^{\dagger}(-{\bm k}) \hat{a}({\bm k}) \hat{a}(-{\bm k}) \rangle$, where $\hat{a}^{\dagger}({\bm k})$ (resp $\hat{a}({\bm k})$) is the creation (resp. annihilation) operator associated with momentum ${\bm k}$.


The experiment starts with the production of weakly-interacting $^4$He$^*$ BECs of a few thousand atoms in a three-dimensional (3D) optical lattice \cite{cayla2018single,carcy2021}. The lattice amplitude is fixed at $V=7.75 \,E_{\mathrm{r}}$, where $E_{\mathrm{r}}/h=h/8md^2$ is the recoil energy, $d=775 \, \mathrm{nm}$ the lattice spacing and $k_d=2 \pi/d$ the associated momentum scale. Using a shallow lattice allows us to enhance interactions and increase the quantum depletion (from 0.2$\%$ to 5$\%$), while remaining in the validity range of the Bogoliubov description. The lattice is then abruptly switched off and the gas let to expand during a long TOF of $296\, \mathrm{ms}$. In a TOF experiment from a lattice, interactions are effectively switched off abruptly \cite{tenart2020two,cayla2018single} and the in-trap many-body wave function is projected onto the momentum basis. Therefore, the measured 3D atom coordinates yield the in-trap momentum $\bm{k}$ of each atom using a ballistic relation. In the reconstructed atom distributions, we can post-select atoms belonging to a specific volume $\Omega_k$ and compute atom-atom correlations over $\Omega_k$. We use this to exclude atoms from the BEC and study only the depletion (see Fig.~\ref{Fig1}b). Finally, statistical averages are obtained from recording about $2,000$ atom distributions (see Methods). To identify pairs, we use the integrated atom-atom correlations,

\begin{equation}
    g_{A}^{(2)} (\delta {\bm k})=\frac{\int_{\Omega_{k}} \langle \hat{a}^{\dagger}({\bm k}) \hat{a}^{\dagger}(\delta {\bm k}-{\bm k}) \hat{a}({\bm k}) \hat{a}(\delta {\bm k}-{\bm k}) \rangle \mathrm{d}{\bm k}}{\int_{\Omega_{k}} \rho({\bm k}) \rho(\delta {\bm k}-{\bm k}) \mathrm{d}\bm{k}},
    \label{Eq:g2}
\end{equation}
where $\rho({\bm k})=\langle \hat{a}^{\dagger} ({\bm k}) \hat{a}({\bm k})  \rangle$. With this definition, a peak located at $\delta {\bm k}={\bm 0}$ signals pairs of atoms at opposite momenta.

\begin{figure*}
    \centering
    \includegraphics[width=1.8\columnwidth]{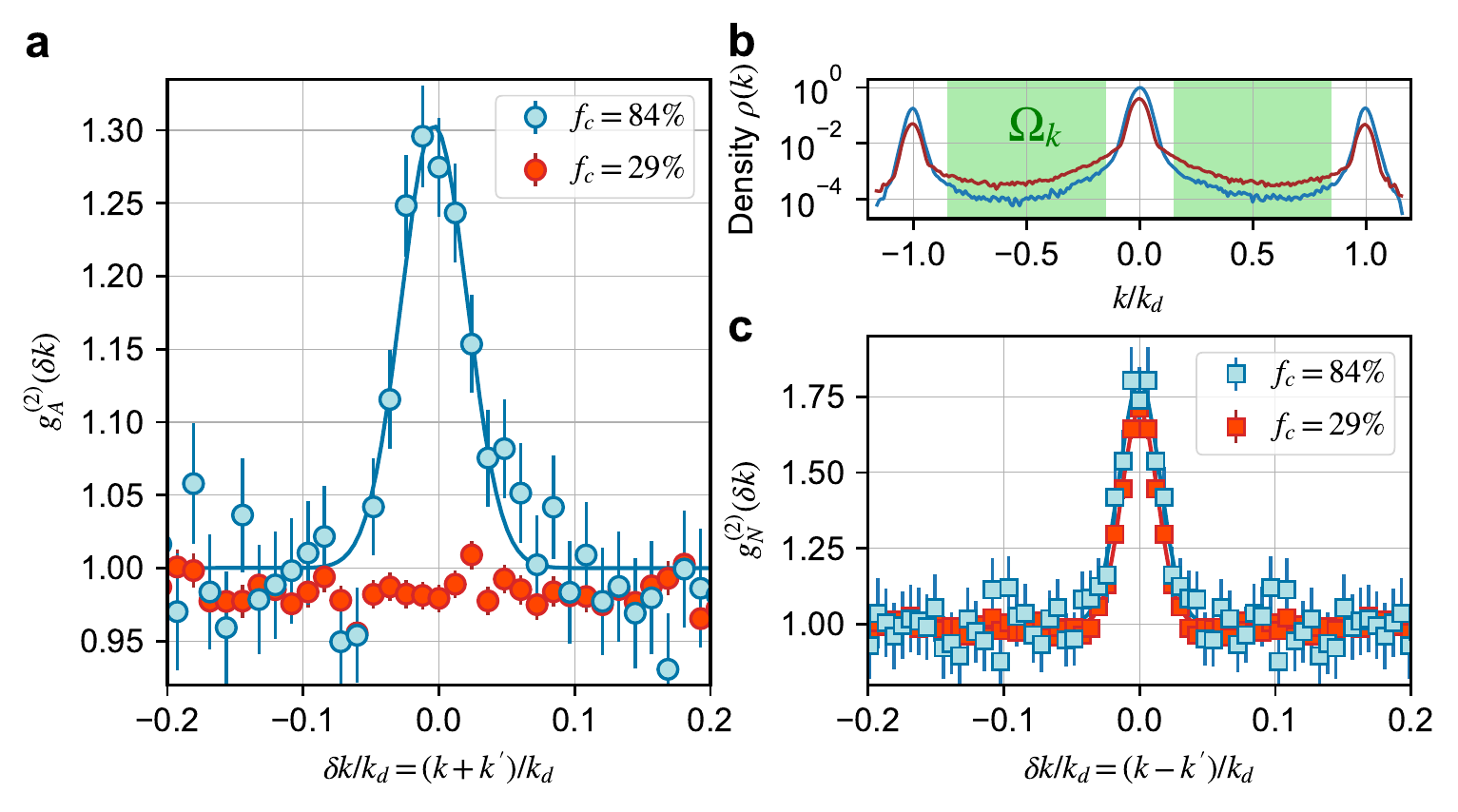}
\caption{Atom-atom correlations in weakly-interacting BECs at two different temperatures. The data for the low-temperature BEC ($f_{c}=84\%$), {\it resp.} for the heated BEC ($f_{c}=29\%$), are depicted in blue, {\it resp.} in red. 
    {\bf (a)}. Anomalous correlations $g_{A}^{(2)}(\delta k)$ at opposite momenta. The ${\bm k}$/$-{\bm k}$ peak disappears as the temperature increases. The RMS width of the peak in the low temperature BEC is $\sigma_{A}=2.4(2) \times 10^{-2} \, k_{d}$.
    {\bf (b)}. 1D cut of the density $\rho(k)$ in semilog scale. The depletion density of the heated BEC is around 4 times larger than that of the low-temperature BEC. It implies a decrease of $g_{A}^{(2)}({\bm 0})-1$ by a factor of around $4^2$ in the heated BEC, since only the denominator in Eq.~\ref{Eq:g2} increases with the thermal depletion. This reduction brings the correlation peak amplitude below the experimental noise. The green shaded area marks the volume over which the correlations are calculated.
    {\bf (c)}. Normal correlations $g_{N}^{(2)}(\delta k)$ for the same datasets and $\Omega_k$. The peak amplitude shows no significant change as the temperature increases. The RMS width is slightly larger for the low-temperature BEC ($\sigma_{N}=1.5(1) \times 10^{-2} \, k_{d}$) than for the heated BEC ($\sigma_{N}=1.3(1) \times 10^{-2} \, k_{d}$). Note that the transverse integration $\Delta k_{\perp}=1.5 \times 10^{-2} \, k_d$ used here reduces the amplitude of the peaks (see Supp. Info.).}
    \label{Fig2}
\end{figure*}

In Fig.~\ref{Fig1}c we present 1D cuts of the pair correlations $g_{A}^{(2)}$ measured in the depletion of lattice BECs, and observe a peak located at $\delta {\bm k}={\bm 0}$. For these data, we find on average about $100$ atoms and $0.5$ atom pairs per shot in $\Omega_{k}$ (see Supp. Info.). A crucial experimental parameter for obtaining this signal is the detection efficiency, which we have recently increased to $0.53(2)$ (see Methods). The observation of atom pairs with opposite momenta in the depletion of equilibrium interacting BECs is a central result of this work. Identifying their origin, however, requires accounting for the effect of temperature.

In our experiment, the temperature $T$ should increase the thermal population of the depletion without contributing to the ${\bm k}$/$-{\bm k}$ correlations. This is because we probe large momenta corresponding to single-particle excitations of the Bogoliubov spectrum (see below). Therefore, when the temperature increases, the number of pairs becomes a negligible fraction of the total depletion, making their detection nearly impossible. This suggests that the ${\bm k}$/$-{\bm k}$ peak vanishes rapidly with temperature, a sensitivity limiting its range of observation but also providing us with a means to confirm its origin. Indeed, an essential aspect of our experiment is the ability to produce BECs in the low-temperature regime $k_B T \ll \mu$ where the thermal depletion ($\sim 10\%$ in Fig.~\ref{Fig1}) is not much greater than the quantum depletion ($\sim 5\%$ in Fig.~\ref{Fig1}). Here, $k_{B}$ is the Boltzmann constant and $\mu$ is the chemical potential. This low-temperature regime, $k_{B}T/\mu \simeq 0.3$ in Fig.~\ref{Fig1}, is accessible in the lattice because it enhances interactions \cite{cayla2020, carcy2021}. 

To study the temperature sensitivity of the ${\bm k}$/$-{\bm k}$ peak, we increase the gas temperature slightly as to maintain a large BEC (see Methods), and repeat the correlation measurement. The two datasets (non-heated and heated) are shown in Fig.~\ref{Fig2}. The increase of temperature translates into an increase of the density $\rho(\bm{k})$, visible in the log-scale plot of Fig.~\ref{Fig2}b. No ${\bm k}$/$-{\bm k}$ peak is visible in the heated BEC, confirming that finite temperature does not contribute to ${\bm k}$/$-{\bm k}$ correlations (see Fig.~\ref{Fig2}a). Our description is further validated by the observation of a ${\bm k}$/$-{\bm k}$ peak of intermediate amplitude at intermediate temperature (see Supp. Info.).

It is also illuminating to contrast the temperature sensitivity of the ${\bm k}$/$-{\bm k}$ correlations with that of local correlations at ${\bm k}'\simeq{\bm k}$. These local correlations reflect bosonic bunching \cite{cayla2020} and are quantified by 
\begin{equation}
    g_{N}^{(2)} (\delta {\bm k})=\frac{\int_{\Omega_{k}} \langle \hat{a}^{\dagger}({\bm k}) \hat{a}^{\dagger}(\delta {\bm k}+{\bm k}) \hat{a}({\bm k}) \hat{a}(\delta {\bm k}+{\bm k}) \rangle \mathrm{d}\bm{k}}{\int_{\Omega_{k}}  \rho({\bm k})  \rho(\delta {\bm k}+{\bm k})  \mathrm{d}\bm{k}},
    \label{Eq:g2N}
\end{equation}
where a peak located at $\delta {\bm k}={\bm 0}$ signals the bunching. In Fig.~\ref{Fig2}c, we plot $g_{N}^{(2)} (\delta {\bm k})$ for both the low-temperature and the heated datasets. We find a similar bunching in both cases. The contrasted behavior of the ${\bm k}$/$-{\bm k}$ and ${\bm k}$/${\bm k}$ correlations with temperature highlights their different nature: the ${\bm k}$/$-{\bm k}$ correlations reflect the quantum correlations present at $T=0$ induced by the atom pairs;  the ${\bm k}$/${\bm k}$ correlations correspond to bosonic bunching revealing thermal (chaotic) statistics, independently of $T$. Having identified the origin of the ${\bm k}$/$-{\bm k}$ correlations, we proceed with a quantitative characterisation to assess their role in the properties of the equilibrium interacting Bose gas.

Our microscopic picture of correlations builds on the Bogoliubov approximation, according to which the Hamiltonian of weakly-interacting bosons is diagonal in the quasi-particle basis \cite{bogoliubov1947}. Therefore, all quantum states have Gaussian statistics and Wick's theorem applies. Correlations can thus be decomposed into two parts, $\langle \hat{a}^{\dagger}({\bm k}) \hat{a}^{\dagger}({\bm k}') \hat{a}({\bm k}) a({\bm k}') \rangle= G^{(2)}_{A}({\bm k},{\bm k}') + G^{(2)}_{N}({\bm k},{\bm k}')$, where $G^{(2)}_{A}({\bm k},{\bm k}')=| \langle \hat{a}^{\dagger}({\bm k}) \hat{a}^{\dagger}({\bm k}') \rangle |^2 $ are the anomalous correlations and $G^{(2)}_{N}({\bm k},{\bm k}')= \rho({\bm k})\rho({\bm k}') + | \langle \hat{a}^{\dagger}({\bm k}) \hat{a}({\bm k}') \rangle |^2$ are the normal correlations. We base our analysis on the fact that the contributions of anomalous and normal correlations are well separated in the momentum space. Indeed, anomalous correlations are present only for ${\bm k}'\simeq-{\bm k}$ \cite{butera2020} and reflected in the momentum-integrated function $g^{(2)}_{A}$, while normal correlations are present only for ${\bm k}'\simeq{\bm k}$, manifesting in $g^{(2)}_{N}$. Furthermore, while the quantum depletion exhibits both anomalous and normal correlations, thermal excitations contribute only to $g^{(2)}_{N}$ here. This is because only thermal excitations with a strong phononic character, \textit{i.e.} at momenta $| k |  \xi \ll 1$ with $\xi$ the BEC healing length, exhibit anomalous correlations \cite{butera2020}. In the probed range of momenta $0.85 \leq | k | \xi \leq 1.15$, this thermal contribution is negligible. To compare Bogoliubov's prediction to the experiment, we fit the bell-shaped correlation peaks with a Gaussian function, from which we extract the amplitudes ($g^{(2)}_A(\bm{0})$ and $g^{(2)}_N(\bm{0})$) and RMS widths ($\sigma_A$ and $\sigma_N$) of anomalous and normal correlations.

\begin{figure}
    \centering
    \includegraphics[width=0.9\columnwidth]{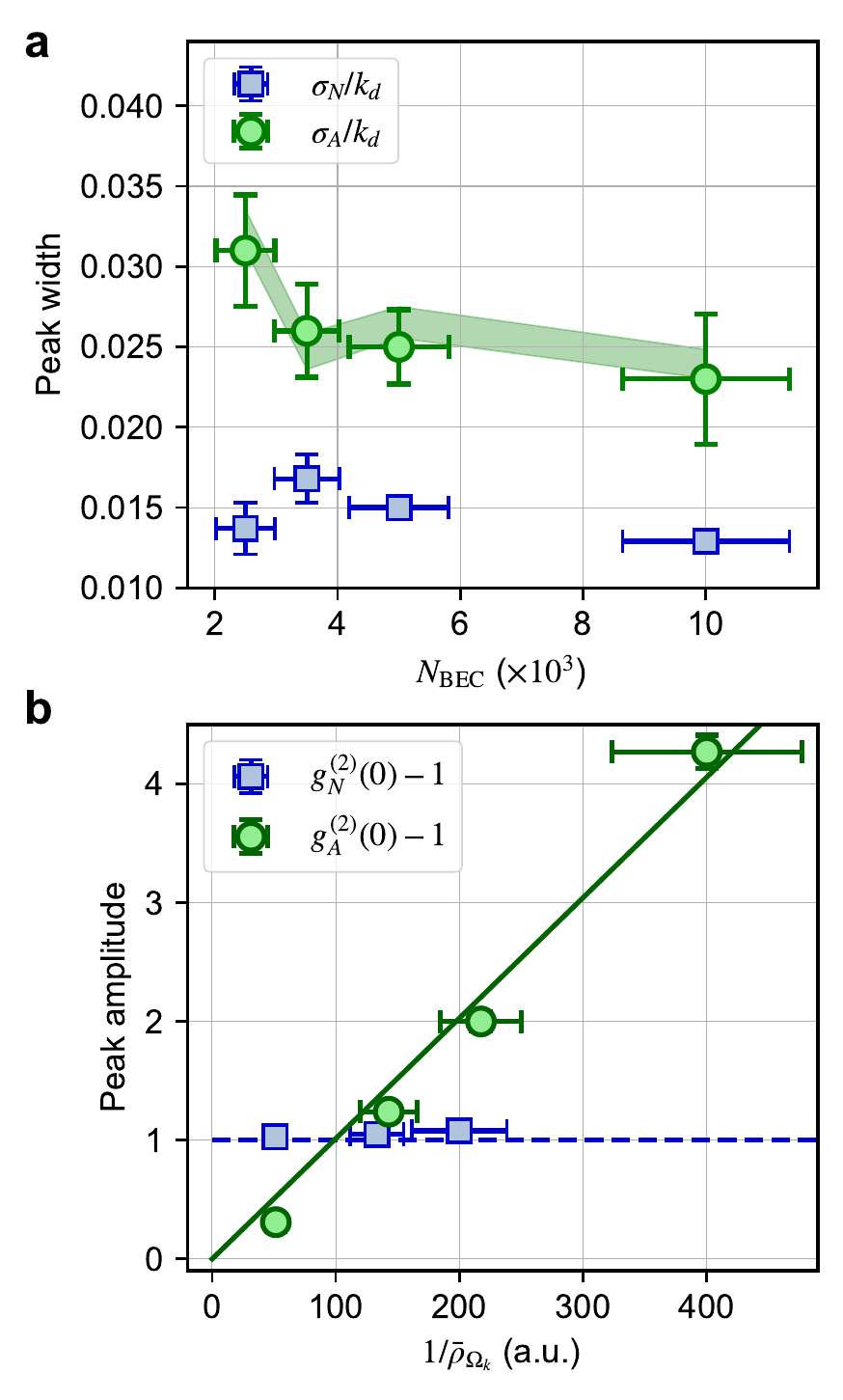}
    \caption{{\bf (a)}. Peak RMS width of the normal correlation $\sigma_{N}$ (blue squares) and of the anomalous correlation $\sigma_{A}$ (green circles) as a function of the BEC atom number $N_{\mathrm{BEC}}$. The green area represents the expected value from the measured BEC width $\sigma_{\rm{BEC}}$ (see Supplementary Information).
    {\bf (b)} Peak amplitude of the normal correlation $g_{N}^{(2)}({\bm 0})-1$ (blue squares) and of the anomalous correlations $g_{A}^{(2)}({\bm 0})-1$ (green circles) as a function of the inverse density $1/\bar{\rho}_{\Omega_{k}}$. While the amplitude of the bunching is perfectly contrasted and constant, $g_{N}^{(2)}({\bm 0})\simeq 2$, the amplitude of the ${\bm k}$/$-{\bm k}$ peak increases linearly with $1/\bar{\rho}_{\Omega_{k}}$. Vertical error bars correspond to the standard deviation of the mean over the three directions of the momentum space. Horizontal error bars correspond to one standard deviation.}
    \label{Fig3}
\end{figure}

The width of the correlation peaks is inversely proportional to the in-trap size of the associated component \cite{cayla2020, butera2020}: $\sigma_N$ is set by the size of the thermal component, while $\sigma_{A}$ is set by the extension of the quantum depletion, limited to that of the BEC by definition. Because the thermal component exceeds the BEC in size, one expects $\sigma_{A} \geq \sigma_{N}$, which is confirmed in all datasets of Fig.~\ref{Fig2}a and Fig.~\ref{Fig3}a. Furthermore, we study how $\sigma_{A}$ varies with the BEC size $L_{\rm BEC}$ by producing low-temperature BECs with varying atom numbers $N$ (from $N=2.5(5)\times10^3$ to $N=10(1)\times10^3$) and by exploiting the fact that the diffraction peaks in $\rho({\bm k})$ have a RMS width $\sigma_{\rm BEC}$ set by $L_{\rm BEC}$. As explained in the Supplementary Information and shown in Fig.-\ref{Fig3}, our measurement is in quantitative agreement with the Bogoliubov prediction $\sigma_A \sim \sigma_{\rm BEC}$ \cite{butera2020} when we account for the effect of small shot-to-shot center-of-mass fluctuations.

\begin{figure}
    \centering
    \includegraphics[width=0.85\columnwidth]{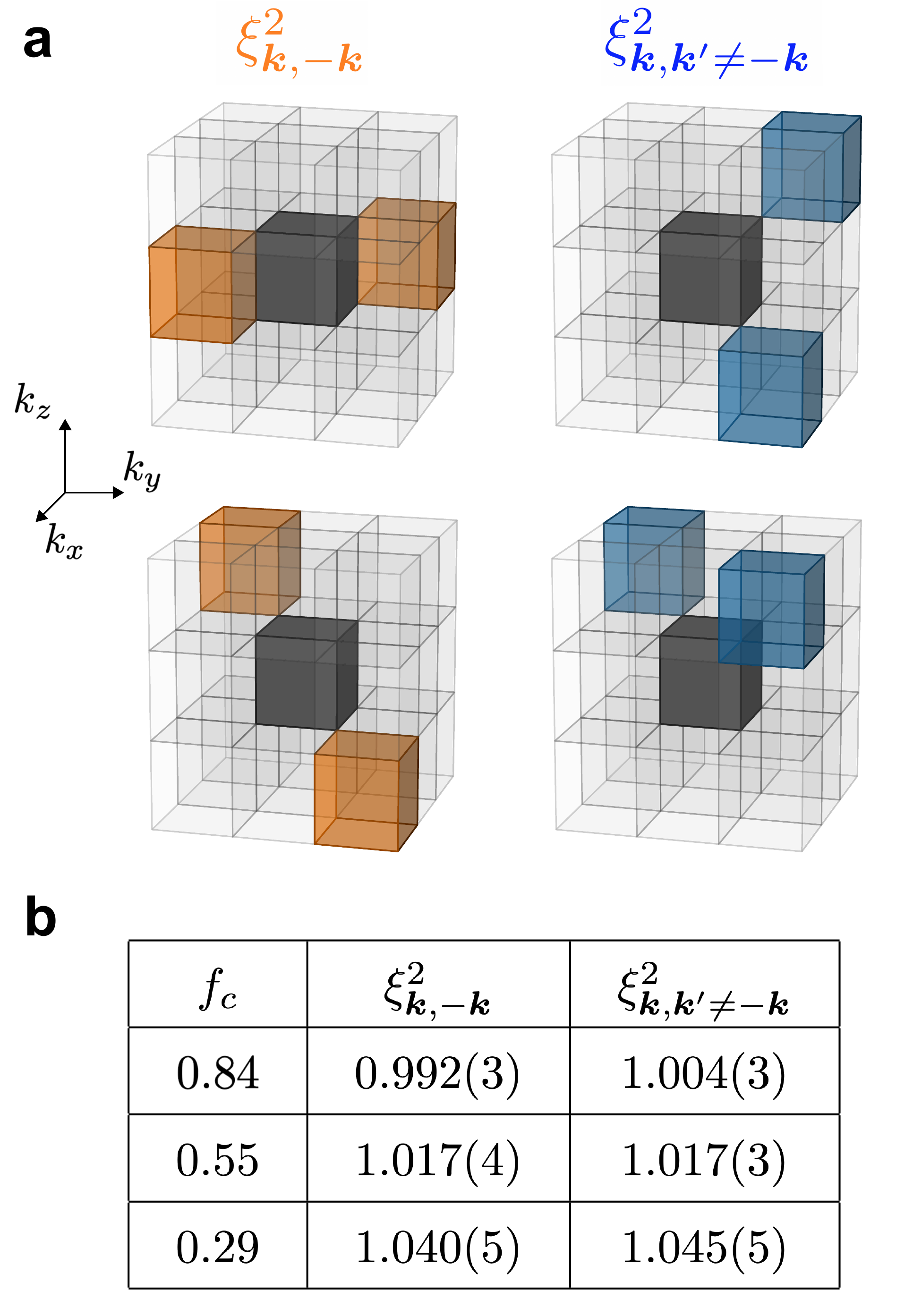}
    \caption{Relative number squeezing. {\bf (a)} Illustration of the momentum-space volumes used to compute the normalized variance $\xi_{\bm{k},\bm{k'}}$ of atom number differences. The BEC is contained in the black voxel centered on $\bm{k}=\bm{0}$. The considered voxels are either located at opposite momenta $\bm{k}$/$-\bm{k}$ (orange) to compute $\xi_{\bm{k},-\bm{k}}$, or at random positions such that $\bm{k'} \neq -\bm{k}$ to compute $\xi_{\bm{k},{\bm k'}\neq-\bm{k}}$ (blue) regions. The voxel size is $(0.3 \, k_d)^3$. The values $\xi_{\bm{k},\bm{k'}}$ are obtained from averaging over all couples of voxels spanning the entire integration volume $\Omega_k$. {\bf (b)} Measured values $\xi_{\bm{k},-\bm{k}}$ for correlated and uncorrelated voxels, and for different values of the condensed fraction $f_{c}$. A small relative number squeezing at opposite momenta, $\xi_{\bm{k},-\bm{k}} < \xi_{\bm{k},{\bm k'}\neq-\bm{k}} \simeq 1$, is observed in the low-temperature data. At higher temperatures, no relative number squeezing is visible. The error interval correspond to the standard error. 
}
\label{Fig4}
\end{figure}

We now discuss the peak amplitudes that reveal the nature of the correlations. As mentioned above, $g_{N}^{(2)}({\bm 0})$ signals bosonic bunching. Theoretically, it is maximally contrasted, $g_{N}^{(2)}({\bm 0})=2$, and unaffected by the temperature because any equilibrium state of the Bogoliubov theory exhibits thermal (chaotic) statistics for ${\bm k}'\simeq{\bm k}$. While this is expected for thermal excitations, it is less obvious for the quantum depletion which is a pure state. In that case, this property results from the destruction of quantum coherences in the measurement of local correlations \cite{cayla2020,butera2020}, as for two-mode squeezed states when only one mode is probed \cite{walls2008}. An analysis of the data of Fig.~\ref{Fig2}c yields $g^{(2)}_{N}({\bm 0}) = 2.05(6)$, confirming that the local statistics are thermal (see Supp. Info.). More importantly, the anomalous amplitude $g^{(2)}_A (\bm{0})$ is expected to increase with the inverse of the density $\rho(\bm{k})$. This is suggested by the Bogoliubov picture of the quantum depletion, based on ${\bm k}$/$-{\bm k}$ pairs: similarly to two-mode squeezed states \cite{loudon1987}, the particle number in one mode $\bm{k}$ (at $T=0$) is equal to the number of pairs, leading to $g^{(2)}_A(\bm{0})-1 \propto 1/\rho(\bm{k})$. To verify this prediction for the measured integrated correlations, we vary the average density $\bar{\rho}_{\Omega_k}=\int_{\Omega_{k}} d{\bm k} \ \rho({\bm k})$ by analysing datasets with different total atom numbers (see Fig.~\ref{Fig3}b). Again, there is a pronounced difference between the normal and anomalous correlations, and we find that $g_{A}^{(2)}({\bm 0})-1$ is indeed inversely proportional to $\bar{\rho}_{\Omega_k}$. Furthermore, the anomalous amplitude reaches values that largely exceed the bunching amplitude. Observing $g^{(2)}_A(\bm{0}) \gg g^{(2)}_N(\bm{0})$ constitutes a violation of the Cauchy-Schwartz inequality for classical fluctuating quantities \cite{walls2008}. It also suggests sub-Poissonian number differences between modes at opposite momenta. A direct calculation on the experimental data confirms this assertion (see Fig.~\ref{Fig4}). We compute the normalized variance of the atom number difference, $\xi_{\bm{k},\bm{k'}}^2=(\langle (N_{\bm{k}} - N_{\bm{k'}})^2 \rangle - \langle N_{\bm{k}} - N_{\bm{k'}} \rangle^2) / (\langle N_{\bm{k}} \rangle + \langle N_{\bm{k'}} \rangle)$ (see Supp. Info.), and measure $\xi_{k,-k}^2=0.992(4)<\xi_{k,k'\neq -k}^2=1.004(4)$ in the low-temperature data (see Fig.~\ref{Fig4}). This number squeezing is small in comparison to that found with discrete (spin) variables \cite{esteve2008squeezing, bucker2011, pezze2018}, being limited by the (uncorrelated) thermal population and the finite detection efficiency. Finally, our observation of $g^{(2)}_{A}(0)\gg g^{(2)}_{N}(0)$ fulfills the Bush-Parentani criterion \cite{busch2014quantum} for certifying entanglement in the continuous variable ${\bm k}$, under the reasonable assumption that $\langle a^{\dagger}({\bm k}) a({-\bm k}) \rangle=0$. However, an experimental confirmation of this assumption is necessary to demonstrate entanglement.

In summary, this work reports the observation of ${\bm k}$/$-{\bm k}$ pairs in the depletion of an interacting Bose gas, and the characterisation of the associated quantum correlations and number squeezing. It also demonstrates that a TOF experiment combined with single-atom detection \cite{bergschneider2019} is a sensitive and quantitative probe of in-trap quantum fluctuations in interacting systems, which can be used to characterise non-Gaussian many-body quantum states \cite{schweigler2017} in the future. A fermionic analogue of this experiment using $^3$He$^*$ to study pairing of momenta would also be of great interest. 

\vspace{1cm}

\paragraph*{Acknowledgements.}
We thank A. Aspect, A. Browaeys, I. Carusotto, H. Cayla and T. Roscilde for their valuable comments on the manuscript, and acknowledge fruitful discussions with the members of the Quantum Gas group at Institut d'Optique. We acknowledge financial support from the LabEx PALM (Grant number ANR-10-LABX-0039), the R\'egion Ile-de-France in the framework of the DIM SIRTEQ, the ``Fondation d'entreprise iXcore pour la Recherche", the Agence Nationale pour la Recherche (Grant number ANR-17-CE30-0020-01). D.C. acknowledges support from the Institut Universitaire de France.


\vspace{1cm}

\begin{center}
{\bf METHODS} 
\end{center}

\paragraph*{Computation of atom-atom correlations} The correlation functions $g_{A}^{(2)} (\delta {\bm k})$ and $g_{N}^{(2)} (\delta {\bm k})$ are computed numerically from the measured atom distributions with a similar two-step procedure. The first step consists in computing the correlations over the momentum-space volume $\Omega_{k}$ in each atom distribution, recorded from a single experimental run, and then averaging the correlations over the set of experimental runs performed under identical conditions. This step computes the numerator in Eq.~\ref{Eq:g2}-\ref{Eq:g2N}, namely $\int_{\Omega_{k}} \langle \hat{a}^{\dagger}({\bm k}) \hat{a}^{\dagger}(\delta {\bm k} \pm {\bm k}) \hat{a}({\bm k}) \hat{a}(\delta {\bm k} \pm {\bm k}) \rangle \mathrm{d}{\bm k}$. The second step aims at evaluating the denominator $\int_{\Omega_{k}}  \rho({\bm k})  \rho(\delta {\bm k}\pm{\bm k})  \mathrm{d}{\bm k}$ in Eq.~\ref{Eq:g2}-\ref{Eq:g2N}  and consists in first merging all the atom distributions used in the first step, and only then computing the correlations. Atom-atom correlations present in a single atom distribution become negligible over the $\sim 2,000$ distributions merged together, so that correlations  in the second step are computed over an ensemble of statistically independent atoms. 

The computed functions $g_{A}^{(2)} (\delta {\bm k})$ and $g_{N}^{(2)} (\delta {\bm k})$ are three-dimensional. When plotting a 1D cut through $g_{A}^{(2)} (\delta {\bm k})$ or $g_{N}^{(2)} (\delta {\bm k})$, a small transverse integration $\Delta k_{\perp}$ is used to improve the signal-to-noise ratio at the expense of reducing the amplitude of the correlation peaks. The amplitudes $g_{A}^{(2)} ({\bm 0})$ and $g_{N}^{(2)} ({\bm 0})$ are obtained from studying the effect of the transverse integration (see Supp. Info.).

\paragraph*{Detection efficiency of the Helium detector} To avoid any disturbance induced by stray magnetic fields over the long free fall ($\sim 45 \, \mathrm{cm}$), the $^4$He$^*$ atoms trapped in the $m_{J}=+1$ sub-state are transferred to the non-magnetic sub-state $m_{J}=0$ after the gas has been released from the optical trap. Any atom that would remain in the $m_J=+1$ state after this step would then be removed by applying a strong magnetic gradient. Here, we use a two-photon Raman transition (detuned by $800 \, \mathrm{MHz}$ from the $2^1$S$_{1} \rightarrow2^3$P$_{0}$ transition) to transfer all the atoms in the $m_{J}=0$ state, so that the detection efficiency per atom is limited only by that of the He$^*$ detector. To optimize the latter, we use funnel-type MCPs (Hamamastu F9142-01 MOD6) whose open-to-air ratio is close to unity ($\sim 0.95$). We have calibrated the total detection efficiency per atom using absorption images as a reference for the atom number and have measured 0.53(2). 

\paragraph*{Heating procedure} The equilibrium lattice BECs in the low-temperature regime are produced by ramping up the lattice amplitude with a rate of $0.3 \, E_{\mathrm{r}}/$ms, and holding the atoms at the final amplitude $V=7.75 \, E_{\mathrm{r}}$ for $5 \, \mathrm{ms}$. At this final lattice amplitude, the ratio of the on-site interaction energy $U$ to the tunnelling energy $J$ is $U/J=5$. We have recently shown that this procedure produces equilibrium states of the Bose-Hubbard Hamiltonian with a low entropy $S/N=0.8(1) \, k_B$ \cite{carcy2021}. From holding the atoms at the final amplitude for a longer duration, the gas is continuously heated over time (attributed to imperfections such as spontaneous emission or mechanical vibrations).  For the heated data shown in Fig.~\ref{Fig2}, we hold the atoms for $500\,$ms at $V=7.75\, E_{\mathrm{r}}$, {\it i.e.} for hundreds of tunnelling times ($225 \times h/J$).


\bibliographystyle{apsrev4-1_our_style.bst}
\bibliography{BogoPairBib.bib}


\cleardoublepage
\begin{widetext}
\begin{center}
\textbf{\large Supplemental Material: Observation of pairs of atoms at opposite momenta in an equilibrium interacting Bose gas}
\end{center}
\end{widetext}

\setcounter{figure}{0} 
\setcounter{equation}{0} 
\setcounter{page}{1}
\renewcommand\theequation{S\arabic{equation}} 
\renewcommand\thefigure{S\arabic{figure}} 
\renewcommand\thepage{S\arabic{page}} 

\section*{Conditions to access the momentum space in a time-of-flight experiment} 

Accessing the momentum-space distribution of trapped bosons in a time-of-flight (TOF) experiment requires {\it (i)} a TOF long enough to reach the far-field regime of expansion \cite{gerbier2008} and {\it (ii)} an expansion which is ballistic (or approximated to be ballistic with an excellent precision). In our experiment, the TOF $t_{\rm TOF}=300~$ms exceeds the far-field time $t_{\rm FF}=4 m L_{\rm BEC}^2/\hbar\simeq 30~$ms, where $L_{\rm BEC}$ is the in-trap radius of the BEC. This ensures that the first condition is fulfilled. As demonstrated in our previous works \cite{cayla2018single, tenart2020two}, the second condition is also fulfilled when interacting bosons are released from a unity-filled 3D lattice. This is mostly due to the fact that the zero-point energy in a lattice site (here equal to $\sim h \times 57.5~$kHz) drives the expansion because it largely exceeds the interaction energy (here equal to $\sim h \times 3.8~$kHz) \cite{gerbier2008, cayla2018single}.

\section*{Integration volume and atom-atom correlations in the BEC} 

As introduced in the main text (see Eq.~1-2), the atom-atom correlations are computed over a volume $\Omega_{k}$ of momentum space that we can choose at will. The main motivation to use a finite volume $\Omega_{k}$ is the ability to investigate the correlations in the BEC and in the depletion of the BEC separately. The type of correlations in the BEC is drastically different from that measured in the depletion presented in the main text. In the BEC, no bunching and no pairs are expected because it is a pure coherent state. In addition, if one were to compute correlations over the entire momentum space, without restricting the computation to a volume $\Omega_{k}$ that excludes the BEC, the correlation signal would be largely dominated by the BEC contribution since the condensed fraction is large ($>80\%$) and the density in the diffraction peaks largely dominates $\rho({\bm k})$. 

In practice, the volume $\Omega_k$ is set to have a cubic symmetry that matches the symmetry of the momentum distribution in a cubic lattice. We remove all atoms with $|k_i| < k_{\rm{min}}$ and $|k_i| > k_{\rm{max}}$ where $k_i$ is the momentum projection along an axis $i=x,y,z$. We use $k_{\rm{min}}=0.15 \, k_d$, corresponding to $\sim 6$ times the RMS width of the BEC peaks, in order to ensure that all condensed atoms have been removed. The high limit is set to $k_{\rm{max}}=0.85 \, k_d$ to exclude higher order peaks and is slightly smaller than the momentum range probed by the MCP. In addition, we can also modify the volume $\Omega_k$ to change the value of the average density $\bar{\rho}_{\Omega_k}$. For the measurement of $g^{(2)}_A(\bm{0})$ in Fig.~3b, we set $k_{\rm{min}}$ to $0.3 \, k_d$ and $k_{\rm{max}}$ to $0.7 \, k_d$ to compute correlations at lower average densities $\bar{\rho}_{\Omega_k}$.

\begin{figure}
    \centering
    \includegraphics[width=0.45\textwidth]{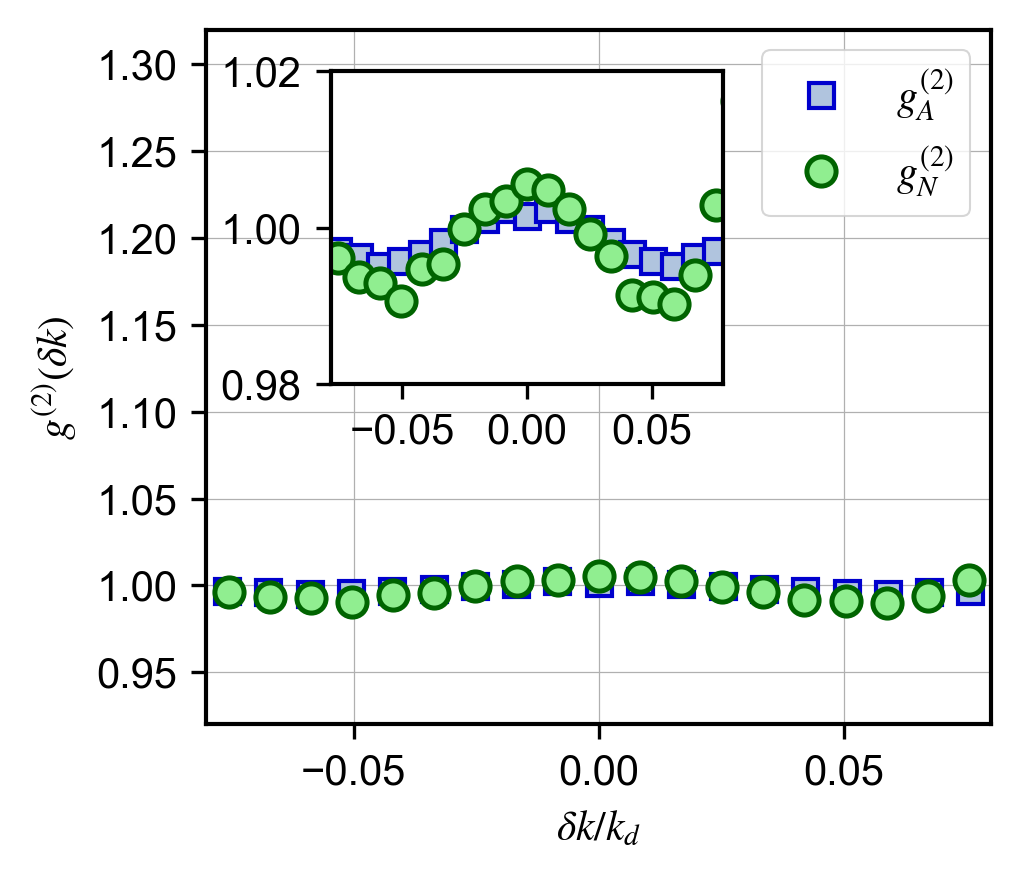}
    \caption{Normal and anomalous correlation functions in the BEC. Note that the vertical scale is of a few percent only. The presence of a small-amplitude modulation is due to normalisation issues induced by shot-to-shot fluctuations of the BEC width $\sigma_{\rm{BEC}}$ in momentum-space. Note that the error bars are smaller than the markers.}
    \label{FigS1}
\end{figure}

To complete the data shown in the main text, we present here the correlations computed for atoms in the BEC. To do so, we post-select atoms with $|k_i| < 0.04 \, k_{d}$ and plot the corresponding correlation functions $g^{(2)}_A$ and $g^{(2)}_N$ in Fig.~\ref{FigS1}. As expected for a coherent state, we do not observe any correlation peak. However, we observe a small modulation at the 1\% level, which is due to shot-to-shot fluctuations of the BEC width $\sigma_{\rm BEC}$. 

\section*{Periodical structure of correlations in the lattice} 

\begin{figure}
    \centering
    \includegraphics[width=0.45\textwidth]{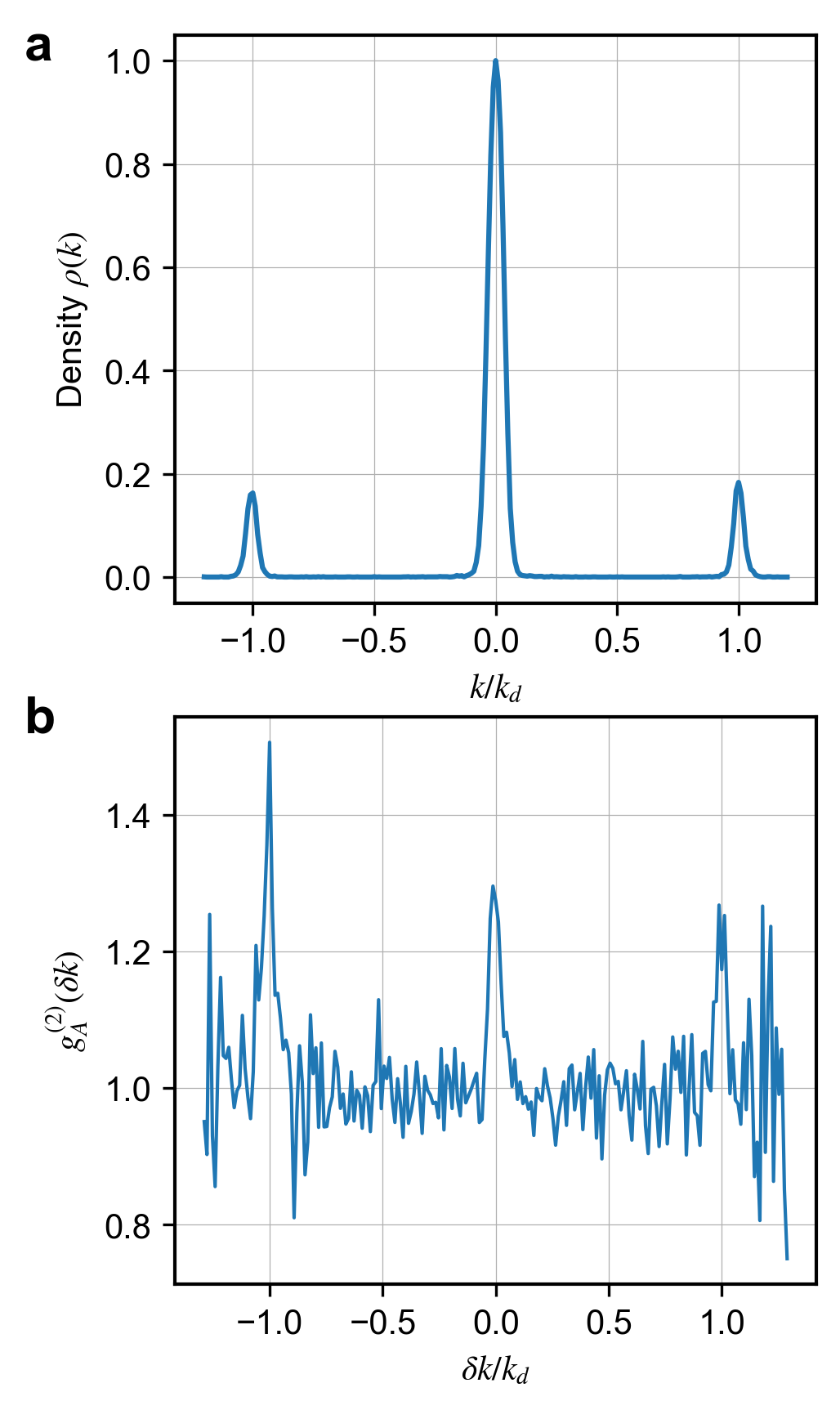}
    \caption{{\bf (a)} 1D cut $\rho(k)$ through the experimental momentum density $\rho({\bm k})$. Diffraction peaks located at $k=0$ and $\pm k_d$ are observed, which signals the phase coherence of the in-trap BEC state. The peaks located at $\pm k_d$ are copies of the one located at  $k=0$, associated with the projection of the in-trap quasi-momentum distribution on the momentum basis through the time-of-flight measurement. {\bf (b)} Anomalous correlation function $g^{(2)}_A(\delta k)$. As for diffraction peaks in the momentum density, we observe additional well-resolved correlation peaks at $\pm k_d$. The transverse integration is $\Delta k_{\perp}=3 \times 10^{-2} \ k_d$.}
    \label{FigS2}
\end{figure}

In a periodical potential like an optical lattice, the momentum is defined in terms of quasi-momentum within the first Brillouin zone. Any physical quantity which is a function of the quasi-momentum is thus periodic when expressed in the momentum basis. In the experiment, we measure the atom distribution in the momentum basis and we expect the various quantities to be periodic. This is the case for the momentum density $\rho({\bm k})$ plotted in Fig.~\ref{FigS2}a, where diffraction peaks separated by $k_d$ are visible. The relative amplitude of these peaks is set by the Fourier transform of the Wannier function of a lattice site. 

Likewise, any correlation function should also be periodic with a period $k_{d}$. Such a periodic structure was previously reported for the normal correlations measured in a Mott insulator  \cite{carcy2019momentum}. Here, we show an example of anomalous correlations $g^{(2)}_A$ (Fig.~\ref{FigS2}b), plotted in the momentum basis. Correlation peaks located at $\delta k=\pm k_d$ are indeed present. Note that the peak amplitudes at $\delta k=\pm k_d$ are expected to be larger than those at $\delta k=0$ as the density $\rho({\bm k})$ probed at larger separation is lower, and the anomalous peak amplitude is inversely proportional to $\rho({\bm k})$. Note that we hardly observe higher order peaks in the normal correlations as their amplitude does not increase at lower densities, resulting in a decrease of the signal-to-noise ratio.

\section*{1D cuts and effect of the transverse integration} 

\begin{figure}
    \centering
    \includegraphics[width=0.5\textwidth]{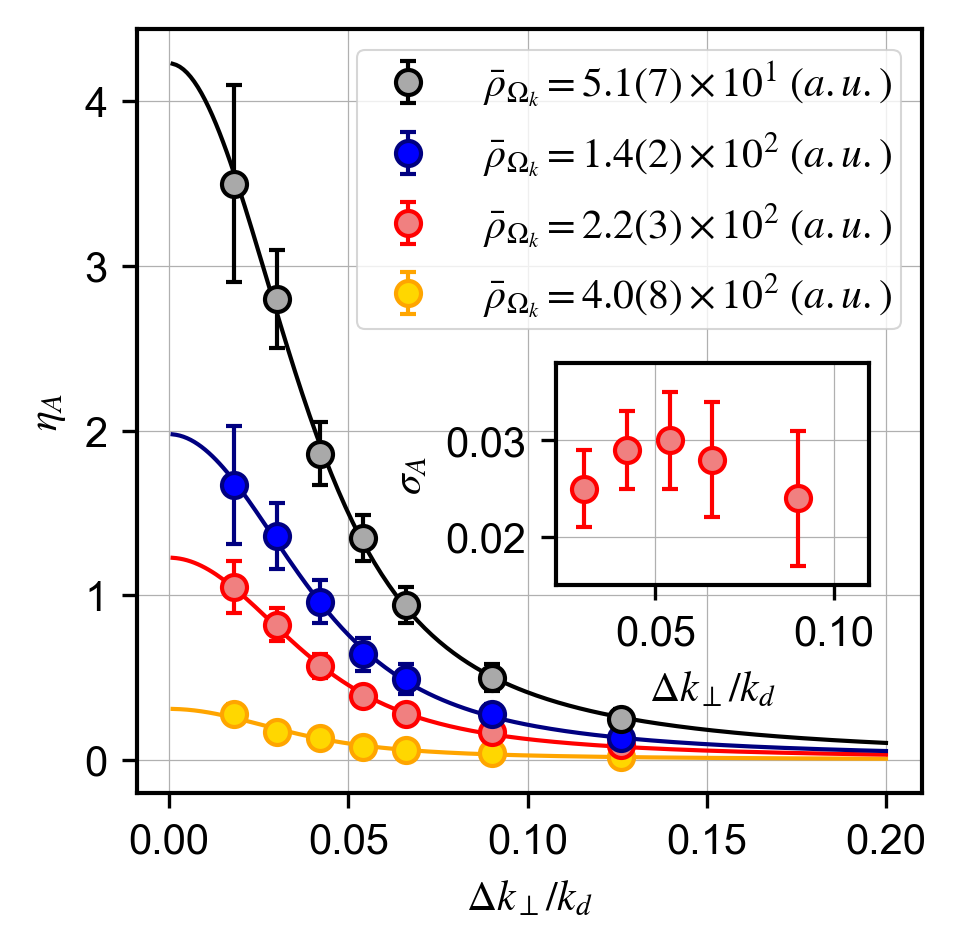}
    \caption{Evolution of the anomalous $\eta_A$ correlation peak amplitude with transverse integration $\Delta k_{\perp}$ for various momentum densities $\bar{\rho}_{\Omega_k}$. The solid lines are fits of the experimental data with Eq.~\ref{EqS1} from which the amplitude $\eta_{0,A}=g^{(2)}_A(\bm{0})$ at vanishing $\Delta k_{\perp}$ is extracted. The inset depicts the evolution of the fitted width $\sigma_A$ with transverse integration. Vertical error bars correspond to the standard deviation of the mean over the three directions of the momentum space.  
}
    \label{FigS3}
\end{figure}

\begin{figure}
    \centering
    \includegraphics[width=0.5\textwidth]{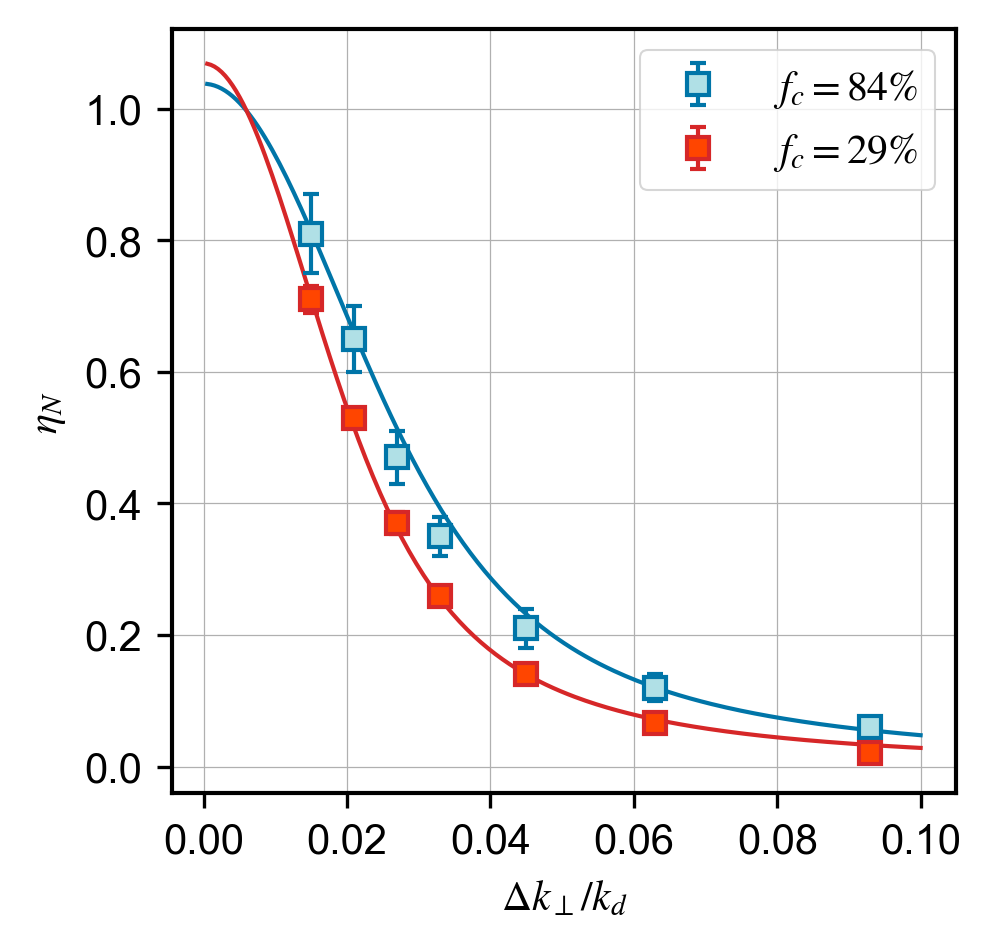}
    \caption{Evolution of the normal $\eta_N$ correlation peak amplitude with transverse integration $\Delta k_{\perp}$ in the low-temperature and heated data. The solid lines are fits with Eq.~\ref{EqS1}. Vertical error bars correspond to the standard deviation of the mean over the three directions of the momentum space. 
}
    \label{Fig_S4}
\end{figure}

In order to plot the 3D correlation functions $g^{(2)}_{A}$ and $g^{(2)}_{N}$, we extract 1D cuts with a small transverse integration $\Delta k_{\perp}$ to increase the signal-to-noise ratio. This transverse integration affects the correlation peaks, since $\Delta k_{\perp}$ is not negligible with respect to the peak widths $\sigma_{A}$ and $\sigma_{N}$. To monitor this effect, we extract the amplitude $\eta$ and the RMS width $\sigma$ of the correlation peaks as a function of the transverse integration $\Delta k_{\perp}$. The longitudinal discretisation is set by the dimension of the histogram bins and chosen to be small enough to ensure proper fitting of the correlation peak widths, while conserving a good signal-to-noise ratio. We choose $\Delta k_{\parallel}=1.2 \times 10^{-2} k_d$ for ${\bm k}$/$-{\bm k}$ correlations and $\Delta k_{\parallel}=6 \times 10^{-3} k_d$ for ${\bm k}$/${\bm k}$ correlations. In Fig.~\ref{FigS3} we plot the results of this analysis of the transverse integration for the case of the anomalous correlations. As shown in the inset, the transverse integration does not have a strong effect on the fitted width $\sigma_A$, with variations contained within the errorbars. This is because the contributions to the correlation of each lattice axis are separable. We then select the value obtained for the lowest transverse integration value providing a satisfactory signal-to-noise ratio. On the other hand, the amplitude $\eta_A$ is strongly affected by the value of $\Delta k_{\perp}$ used to plot the correlation peak. Because the correlation peaks are bell-shaped, close to isotropic and well fitted by a 3D Gaussian function, this dependency takes the following form:

\begin{equation}
    \eta_{A}(\Delta k_{\perp})= \eta_{0,A} \times \frac{2 \pi \sigma_A^2}{(2\Delta k_{\perp})^2}\left[\rm{erf} \left(\frac{\Delta k_{\perp}}{\sqrt{2}\sigma_A} \right)\right]^2
    \label{EqS1}
\end{equation}

where $\rm{erf}(x)=\frac{2}{\pi} \int_0^x \exp (-y^2) \rm{d}y$. From fitting the data shown in Fig.~\ref{FigS3} with this dependency where both $\eta_{0,A}$ and $\sigma_A$ are fitting parameters, we extract the amplitude at vanishing integration $\eta_{0,A}=g^{(2)}_{A}({\bm 0})$. In addition, we ensure that the fitted value of $\sigma_A$ is compatible with the one fitted directly to the correlation peak as explained previously. We proceed similarly to extract the properties of the normal correlations, an approach that we recently used to characterise Mott insulators \cite{cayla2020}. This is illustrated in Fig.~\ref{Fig_S4} where the analysis is conducted for the normal correlations of the two data sets in Fig.~2c. We obtain $g^{(2)}_N(\bm{0})=2.05(6)$ and $g^{(2)}_N(\bm{0})=2.09(5)$ for the high and low condensed fraction data sets respectively, consistently with the expected $g^{(2)}_N(\bm{0})=2$ for bosonic bunching.

Examples of 1D cuts with a fixed transverse integration are plotted in Fig.~1 of the main text. In Fig.~2 we average the cuts in the three directions of the lattice to obtain a single plot. In a similar fashion, the amplitude and width values for a given value $\Delta k_{\perp}$ of transverse integration are obtained by averaging the fitted values on the three different 1D cuts.
 
\section*{Total number of pairs}

We describe here how we evaluate the number $N_{\rm{pairs}}(\Omega_k)$ of atom pairs detected in the volume $\Omega_k$. We first increase the bin size to count every atom pair in the central voxel of the 3D histogram used to compute correlations. When calculating the histograms shot by shot, both ``true coincidences'' linked to the presence of a ${\bm k}$/$-{\bm k}$ pair as well ``accidental coincidences'' related to the density distribution of the atoms are summed up. To evaluate the number of accidental coincidences associated with the measured densities $\rho(\bm{k})$, we merge all atom distributions to obtain an ensemble of statistically independent atoms with the same density $\rho(\bm{k})$. Then we can substract this number of accidental coincidences from the number of total coincidences to obtain the number of pairs $N_{\rm{pairs}}$.

We have quantitatively tested the validity of this procedure by studying a different experimental configuration with lattice Bose gases with a large number of atoms ($N\sim 10^5$). Under these conditions, the population of the diffraction peaks during the TOF is large enough for elastic collisions to take place. As a result, collisional spheres are observed between the diffraction peaks, for which we can accurately estimate the number of pairs, as illustrated in a previous work \cite{tenart2020two}. We thus compare this prediction with the procedure described above to estimate $N_{\rm{pairs}}(\Omega_k)$, finding an excellent agreement when accounting for the finite detection efficiency.  This comparison therefore also confirms the measured value of the detection efficiency described in the Methods.

Note that we are unable to compare the measured number of pairs to a theoretical prediction. This is because a quantitative description in the regime of the experiment is currently difficult. On the one hand,  a quantitative description within the Bogoliubov approximation requires extending the work \cite{butera2020} performed in 1D trapped systems to the 3D lattice geometry we study in this work.  On the other hand, ab-initio approaches for the 3D Bose-Hubbard model such as Quantum Monte-Carlo simulations have been limited to calculate densities, although they could also be used to calculate two-body correlations in principle.

\section*{Effect of temperature on the pair correlations}

\begin{figure}
    \centering
    \includegraphics[width=0.45\textwidth]{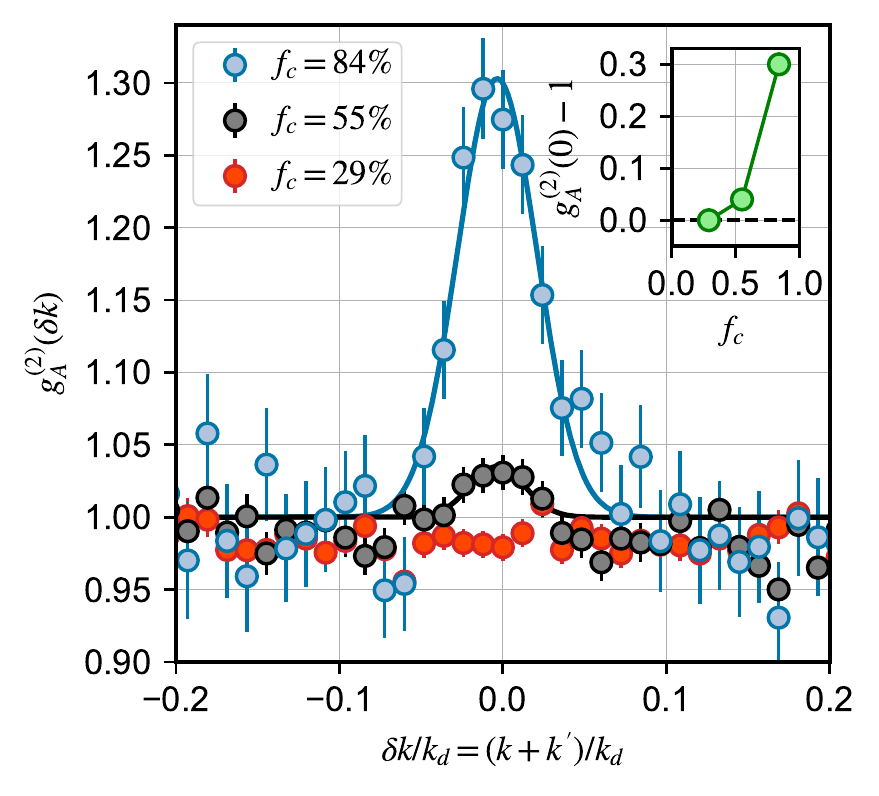}
    \caption{Anomalous correlation function for data sets with different temperatures and condensed fractions. The amplitude of the $\bm{k}$/$-\bm{k}$ correlation signal is progressively lost as the temperature rises and the condensed fraction diminishes. Inset: amplitude of the correlation peak $g^{(2}_{A}({\bm 0})$ as a function of the condensed fraction $f_c$.}
    \label{FigS5}
\end{figure}

To complement the discussion of the evolution of the ${\bm k}$/$-{\bm k}$ correlation peak with respect to temperature, we plot in Fig.~\ref{FigS5} the anomalous correlation function $g^{(2)}_A$ for an additional dataset with $f_c=55 \%$. This third dataset is obtained with a holding time of $200\,$ms, corresponding to $90 \times h/J$. The two datasets shown in the main text are also plotted. At this intermediate temperature, the ${\bm k}$/$-{\bm k}$ signal is visible but its amplitude is significantly lower than that of the low-temperature BEC. This further illustrates how the ${\bm k}$/$-{\bm k}$ correlation peak is progressively lost as the temperature rises. We note that the noise is reduced as the temperature increases since the total depletion is larger.  

\section*{Effect of shot-to-shot center-of-mass fluctuations} 

Here, we discuss the effects of shot-to-shot center-of-mass fluctuations on the measured BEC width $\sigma_{\rm BEC}$ and the anomalous width $\sigma_{A}$. We consider center-of-mass displacements of a small amplitude, at the level of 1\% of the lattice recoil $k_{d}$, larger than the outstanding resolution of the He$^*$ detector (RMS width $0.003k_{d}$ \cite{cayla2020}). The fitting procedures we have tested have failed to correctly locate the center-of-mass on a single shot with such a precision. To model these center-of-mass fluctuations, we assume that the center-of-mass displacement from one shot to another has a Gaussian distribution, centered on ${\bm k}={\bm 0}$ and of RMS width $\Delta k_{\rm{com}}$. The effect of the center-of-mass fluctuations is to convolve the bare experimental data with this distribution. When $\Delta k_{\rm{com}}$ is sufficiently large, one expects both $\sigma_{\rm BEC}$ and $\sigma_{A}$ to increase. Note, however, that the measurement of the local correlations is not affected by the center-of-mass fluctuations, {\it i.e.} $\sigma_{N}$ is left unchanged: within a given shot where two-body correlations are calculated, the center-of-mass fluctuations simply manifest as a global displacement of all the atoms of this shot, say by a quantity $\bm{k}_{\mathrm{COM}}$. As a result, the momentum difference $\delta \bm{k} = \bm{k}_1 - \bm{k}_2$ between two atoms of this shot is not affected by the center-of-mass fluctuations, $(\bm{k}_1+\bm{k}_{\mathrm{COM}}) - (\bm{k}_2+\bm{k}_{\mathrm{COM}})=\bm{k}_1 - \bm{k}_2$.

The momentum density $\rho({\bm k})$ of a 3D harmonically-trapped BEC in the Thomas-Fermi regime is well approximated by a Gaussian shape of RMS width $\sigma_{\rm BEC,0} \simeq 1.6/L_{\rm BEC}$. In the presence of center-of-mass fluctuations, the measured momentum density results from the convolution with the distribution of center-of-mass displacements and has a RMS width $\sigma_{\rm{BEC}}=\sqrt{\sigma_{\rm{BEC},0}^2+\Delta k_{\rm{com}}^2}$. For the lattice configuration explored in the experiment, we evaluate $L_{\rm BEC}$ using the Gutzwiller variational approach \cite{jaksch1998cold}. For instance, we  find $\sigma_{\rm BEC,0} \simeq 1.7 \times 10^{-2} \ k_d$ for a total atom number $N= 5 \times 10^3$. From the measured value $\sigma_{\rm BEC} = 2.00(4) \times 10^{-2} \ k_d$, we obtain a quantitative estimate of the RMS width $\Delta k_{\rm{com}}=1.05(2) \times 10^{-2} \ k_d$ of the center-of-mass fluctuations. We repeat this procedure to evaluate $\Delta k_{\rm{com}}$ for each data set.

Likewise, the peak of the anomalous correlations is enlarged by the center-of-mass fluctuations, with respect to its value $\sigma_{A,0}$ expected in the absence of such fluctuations. For a ${\bm k}$/$-{\bm k}$ pair of atoms, a center-of-mass displacement of amplitude $\mathrm{d}{\bm k}$ results in a momentum difference $\delta {\bm k}=2 \mathrm{d}{\bm k}$. The effect of the center-of-mass displacement on $\sigma_{A}$ is thus twice as large as that on $\sigma_{\rm BEC}$, and we obtain $\sigma_A = \sqrt{\sigma_{A,0}^2+4\Delta k_{\rm{com}}^2}$. From numerical calculations for interacting trapped Bose gases in the Bogoliubov approximation \cite{butera2020}, the bare anomalous width is predicted to be equal to $\sigma_{A,0} \simeq 0.94 \ \sigma_{\rm BEC,0}$. In the presence of a lattice, this relation is not modified since the size $L_{\rm BEC}$ does not change when the ratio $\mu / \hbar \omega$ is fixed ($m \omega^2=m^* {\omega^*}^2$ with $m^*$ the effective mass in the lattice and $\omega^*$ the corresponding effective frequency). Combining this relation with the evaluated values of $\Delta k_{\rm{com}}$, we obtain a corrected estimate of $\sigma_A$ represented as the green area in Fig.~3a. The uncertainty is given by the uncertainty on the measurement of $\sigma_{\rm BEC}$ and the uncertainty on the determination of $L_{\rm BEC}$ caused by fluctuations of the total atom number.

\section*{Relative number squeezing}
We detail in this section the procedure used to compute the normalized atom number difference variance $\xi_{\bm{k},\bm{k'}}^2=(\langle (N_{\bm{k}} - N_{\bm{k'}})^2 \rangle - \langle N_{\bm{k}} - N_{\bm{k'}} \rangle^2) / (\langle N_{\bm{k}} \rangle + \langle N_{\bm{k'}} \rangle)$ and to detect relative number squeezing. With the previous expression, if $N_{\bm{k}}$ and $N_{\bm{k}'}$ are independent variables with Poissonian statistics, their difference is Poissonian as well and $\xi^2=1$. Measuring $\xi^2<1$ then signals sub-Poissonian statistics for the number difference associated with relative number squeezing. For each experimental run, we record the number of detected atoms in voxels of size $(0.3 \, k_d)^3$ paving the integration volume $\Omega_k$. The size of the voxels is chosen so that each one contains a sufficient number of detected atoms. The average numbers of detected atoms per shot are $\sim 100$, $\sim 240$ and $\sim 360$ for the data sets with $f_c=84\%$, $f_c=55\%$ and $f_c=29\%$ respectively. The statistical averages in the definition of $\xi^2$ are evaluated on the $\sim 2000$ available experimental shots. We compute $\xi^2$ between ${\bm k}$/$-{\bm k}$ voxels for which we expect relative number squeezing, and for uncorrelated voxels for which we expect $\xi^2=1$. The final value of $\xi^2$ is obtained by averaging over 62 couples of voxels (either located at ${\bm k'}=-{\bm k}$ or at ${\bm k'} \neq -{\bm k}$). The uncertainty is defined as the standard error over the 62 measured values of $\xi^2$. Note that we show the bare values of $\xi^2$ without removing the contributions from the finite detection noise and total atom number fluctuations.

As shown in the main text, we observe that $\xi^2$ increases above 1 for the lower condensed fraction data. This can be explained by fluctuations of the total atom number adding a contribution to the number difference variance proportional to $\langle N_k \rangle^2$, contrary to the Poisson law where the variance is proportional to $\langle N_k \rangle$. As a result, when $\langle N_k \rangle$ is small, {\it i.e.} for high condensed fractions, total atom number fluctuations can be neglected, while this is not the case when $\langle N_k \rangle$ increases, {\it i.e.} for low condensed fractions.


\end{document}